\documentstyle[12pt]{article}

\title{\bf On singularities of the mixed state phase }

\vspace{60mm}

\author{\bf Rajendra~Bhandari}

\date{ }

\begin{document}

\maketitle
\vspace{40mm}
\begin{center}
\begin{tabular}{ll}
            & Raman Research Institute, \\
            & Bangalore 560 080, India. \\
            & email: bhandari@rri.res.in\\
\end{tabular}
\end{center}
\vspace{40mm}
-----------------------------------------------------------------------\\
MS version of 13 August 2001.

\newpage

\par In an interesting contribution to the physics of interference 
of particles with  internal degrees of freedom, Sj\"{o}qvist et.al.
\cite{rhoui} have shown that if a beam of such particles, 
in a mixed state with a density matrix ${\rho}_0$,
is split in a Mach-Zhender interferometer and a unitary transformation 
$U_i$ acts on the space of N internal states in one of the two paths, 
a phase difference given by arg Tr($U_i$${\rho}_0$) is introduced 
between the two beams. Here ``phase difference" is defined as 
the shift in the maximum of 
the interference pattern. For pure states $\mid{\psi}_0>$ 
this quantity reduces to the phase shift 
arg($<{\psi}_0 {\mid U_i \mid} {\psi}_0>$) defined by 
Pancharatnam \cite{panch}.
The purpose of this note is to point out that the definition of the 
mixed state phase as in ref.\cite{rhoui} breaks down at points in the 
parameter space where ${\mid Tr(U_i{\rho}_0)\mid}$=0. Eqn.(8) 
in the paper shows that at such points the interference pattern has 
zero contrast. Such points constitute singularities of 
phase involving discontinuous phase jumps. 
In the pure state case, such singularities have been 
demonstrated theoretically and experimentally for N=2 using the two  
polarization states of light \cite{rbjumps,rbdirac,4pism,rbreview}. 
The mixed state phase involves singularities in new kinds of parameter spaces 
which include variables representing decoherence of quantum states. 
We show that the singularities make interpretation of even simple 
interference experiments like the ones analyzed in ref. \cite{rhoui},
in terms of the proposed phase, nontrivial.

Consider  a beam of spin-1/2 particles in a mixed state with a 
density matrix ${\rho}_0$ = diag$[(1+r)/2 , (1-r)/2]$ in the basis 
of $\mid z \pm>$; $\mid z \pm>$ being the eigenstates of  
z-component of the spin and $r$  the degree of 
polarization. The beam is split symmetrically in an interferometer 
and, to keep things simple,
a unitary transformation $U_i$, diagonal in the $\mid z \pm>$ basis, 
is applied in one of the two paths by means of a variable magnetic field 
along $\hat z$. This results in the $\mid z \pm>$ states acquiring a 
phase factor $e^{\pm i \delta}$. The mixed state phase $\phi$ 
is given by 
$ Tr(U_i{\rho}_0)$ = $[cos \delta + i~ r~ sin \delta]$ = $c~e^{i\phi}$.
In the plane of the parameters $(r, \delta)$,  phase 
singularities occur at the points ($0, (j+1/2)\pi$), 
where $c$ vanishes; $j$ being any integer. 
The interference pattern for the unpolarized case ($r=0$) can be 
looked upon as a 
superposition of two cosine intensity patterns with equal 
amplitude for the two spin eigenstates, moving in the opposite 
directions as $\delta$ is varied \cite{4pism}. When $\delta=(j+1/2)\pi$, 
the relative phase shift of the two patterns equals $(2j+1)\pi$ and 
the superposition yields uniform illumination, making the phase 
indeterminate. A counterclockwise (clockwise) circuit 
in the $(r, \delta)$ plane around 
any one of these singularities yields a value 
$2\pi$ (-$2\pi$) for the total phase change $\int d\phi$.

Fig.1 shows a set of typical phase shifts acquired 
by a mixed state as a function of $\delta$. In each case the 
phase shift equals $\pi$ in magnitude for a variation of 
$\delta$ through $\pi$, is linear for pure states i.e. for 
$r$=$\pm$1 (curves A and B), is highly nonlinear near the 
singularity at ($r$=0, $\delta = \pi/2$) (curves C and D), and 
changes sign with the sign of $r$ ($d\phi/d\delta$ has the same sign 
as $r$). For $r$=0, the phase shift 
is indeterminate, contrary to the claim in  ref.\cite{rhoui} 
of its being equal to $\pi$. 
The neutron interferometer experiments cited therein, done with 
unpolarized neutrons ($r$=0), do not measure the {\it phase shifts} as shown 
in fig.1. These can however be measured in an interference 
experiment with neutrons or with polarized light if $r$
can be varied and phase shifts  measured in the experiment. 
The  basic topological effect namely a  
$2n\pi$ phase change (n being integer) resulting from a 
circuit around one or more 
singularities can in fact be measured without a close 
approach to the singularities, making experiments easier. 
In an experiment with spin$>$1/2, even a circuit around 
a single singularity could 
result in a $2n\pi$ phase shift with $n>1$, yielding information about the 
spin quantum number. A modulo $2\pi$ description 
of the phase, as in ref.\cite{rhoui} and in early work on the 
geometric phase, thus misses an 
essential aspect of the physics in the problem.\\

{\bf Figure Caption:}

{\bf FIG.1:} Computed phase shift $\int d\phi$ as a 
function of applied magnetic field for 
(A) $r$=1, (B) $r$=-1, (C) $r$=.001 and (D) $r$=-.001.\\

 
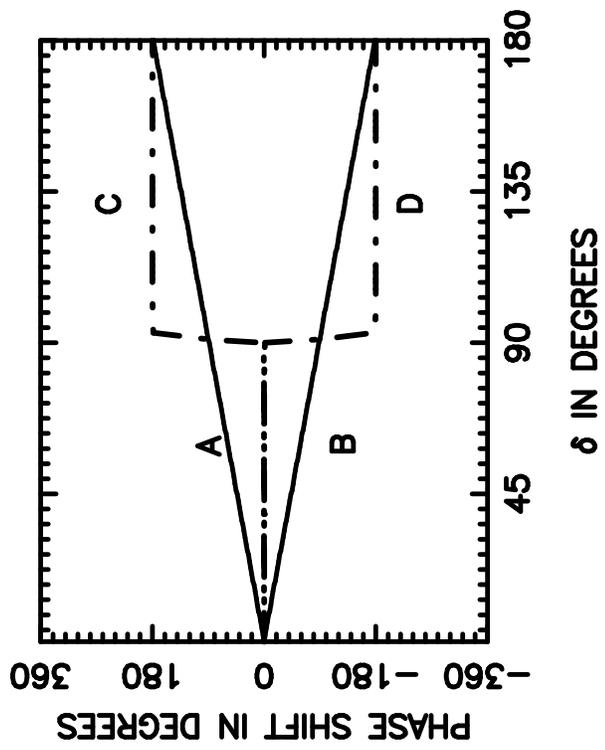
\end{document}